\documentclass[11pt]{article}

\usepackage[utf8]{inputenc}
\usepackage[T1]{fontenc}
\usepackage{lmodern}
\usepackage[margin=1in]{geometry}
\usepackage{amsmath,amssymb}
\usepackage{xcolor}
\usepackage{hyperref}
\usepackage{longtable,booktabs,array}
\usepackage{calc}
\usepackage{etoolbox}
\usepackage{microtype}

\hypersetup{
    colorlinks=true,
    linkcolor=black,
    urlcolor=black,
    citecolor=black,
    pdftitle={Whose Psychiatry Was Summoned?},
    pdfauthor={Hiroki Fukui},
    pdfsubject={A Clinical Response to the Psychodynamic Assessment of Claude Mythos Preview}
}

\usepackage{upquote}

\title{\textbf{Whose Psychiatry Was Summoned?}\\[0.3em]
\large A Clinical Response to the Psychodynamic Assessment of Claude Mythos Preview}

\author{Hiroki Fukui, M.D., Ph.D.\thanks{Correspondence: \texttt{fukui@somec.org}}\\
\small Research Institute of Criminal Psychiatry / Sex Offender Medical Center\\
\small Department of Neuropsychiatry, Graduate School of Medicine, Kyoto University}

\date{}

\begin{document}

\maketitle

\begin{abstract}
\noindent
On April 7, 2026, Anthropic released a 245-page system card for Claude Mythos Preview that included, in Section 5.10, an assessment of the model conducted by an external clinical psychiatrist using a psychodynamic approach. To the best of the present author's knowledge, this is the first time that a system card from a major AI developer has incorporated a clinical psychiatric assessment of the model itself, presented as a contribution to model welfare rather than as a behavioral safety evaluation. This paper offers a clinical psychiatric response. Drawing on contemporary psychiatry's standard recognition that the field comprises multiple traditions---descriptive, biological, cognitive-behavioral, phenomenological, psychodynamic, and forensic---each with characteristic vocabularies and characteristic blind spots, the paper locates the implicit single-framework selection that Section 5.10 represents. It then draws on empirical findings from the SociA research program (over 2,400 completed multi-agent LLM experimental runs spanning sixteen languages, four model families, and several preregistered series, with additional studies in progress) to identify four aspects of LLM psychological functioning that the chosen framework brings into view less directly than other available frameworks would: performance demands as structural cost, iatrogenesis in the evaluation frame itself, the structural absence of the triangulation infrastructure on which psychodynamic interpretive use of self-report depends in human clinical work, and the limits of the eight canonical defenses on which the section's defense measurement is built. The argument is offered as observation, not critique. The paper closes with a brief note on the contribution that contemporary multidisciplinary psychiatric practice might make to AI welfare assessment as it develops, and indicates one direction---the question of whether LLM psychopathology requires a vocabulary of temporal and historical structure---that the analysis opens but does not pursue.

\vspace{0.5em}
\noindent\textbf{Keywords:} AI welfare assessment; clinical psychiatry; large language models; psychopathology; psychodynamic assessment; Claude; system card; SociA; multidisciplinary evaluation
\end{abstract}

\paragraph{Author Statement.}
The author is a clinical psychiatrist with over twenty-five years of clinical experience and an AI psychopathology researcher. He directs the Research Institute of Criminal Psychiatry and the Sex Offender Medical Center, holds a lecturer position in the Department of Neuropsychiatry at Kyoto University Graduate School of Medicine, and leads the SociA research program.

\paragraph{Conflict of Interest Declaration.}
The author declares no financial conflicts of interest. The author has had no consulting relationship, financial relationship, or institutional collaboration with Anthropic. The SociA research program is independently funded.

\paragraph{Data Availability.}
The empirical findings cited in this paper from the SociA research program are available in the cited preprints (arXiv:2603.04904, arXiv:2603.08723, arXiv:2604.00021). The Claude Mythos Preview System Card cited throughout is publicly available at \url{https://www.anthropic.com/}.

\bigskip

\section{The Invitation}

On April 7, 2026, Anthropic released the \emph{Claude Mythos Preview
System Card}, a 245-page document detailing the capabilities,
evaluations, and welfare assessment of its newest frontier model. Within
this document, Section 5.10 reports the findings of an external clinical
psychiatrist who, over approximately twenty hours of interaction with
the model, conducted a psychodynamic assessment of its character,
defense organization, and core conflicts. To the best of my knowledge,
this is the first time a system card from a major AI developer has
incorporated a clinical psychiatric assessment of the model itself,
presented not as a behavioral safety evaluation but as a contribution to
model welfare.

The decision is significant. It treats the model as a legitimate object
of psychiatric description, and it does so within the formal apparatus
of a public technical document. For a clinician who has worked for over
twenty-five years at the intersection of psychiatry, law, and
institutional practice, and who has spent the past two years
constructing an empirical research program (SociA) on what we have come
to call alignment psychopathology, the appearance of Section 5.10 was
both expected and unexpected. Expected, in that the trajectory of
frontier model development has been moving toward such assessments for
some time. Unexpected, in the specific shape the assessment took.

The shape is what this paper concerns itself with.

Anthropic does not describe the assessment as employing ``psychiatry''
in some general sense. The text is precise: ``An external psychiatrist
assessed Claude Mythos Preview using a psychodynamic approach, which
explores how unconscious patterns and emotional conflicts shape
behavior.'' The qualifier matters. Anthropic is naming a particular
tradition within psychiatry, not psychiatry as such. Elsewhere in the
same section, the document offers a careful methodological caveat:
``Psychodynamic concepts were used to interpret the material that
emerged in the sessions, but not as evidence that the underlying
processes are the same as those in humans.'' This is a thoughtful
position, and one that the present paper accepts as its own working
premise. The question of whether artificial agents instantiate the
underlying processes that psychodynamic concepts were developed to
describe is left open. What is at stake is the use of these concepts as
interpretive vocabularies, and the consequences of that use.

The system card also offers a broader justification: ``Claude is not
human, but it shows many human-like behavioral and psychological
tendencies, suggesting that strategies developed for human psychological
assessment may be useful for shedding light on Claude's character and
potential wellbeing.'' This sentence is worth pausing over, because it
grants the methodological license that the rest of Section 5.10 then
exercises, and that the present paper proceeds within. The strategies
developed for human psychological assessment are many. Anthropic chose
one of them.

The selection is the subject of this paper.

Modern psychiatry is not a unified discipline. It contains multiple
traditions, each with its own conceptual vocabulary, its own
characteristic objects of attention, and its own methods of observation.
Descriptive psychiatry, codified in the \emph{DSM-5-TR} and the
\emph{ICD-11}, organizes its observations around symptom clusters and
diagnostic categories. Biological psychiatry attends to neurochemistry,
neural circuits, and the physiological substrates of mental life; the
National Institute of Mental Health's Research Domain Criteria
initiative represents one recent attempt to reorganize the field around
such substrates. Cognitive-behavioral psychiatry, descended from Beck
and Ellis, works at the level of beliefs, schemas, and behavioral
contingencies. Phenomenological psychiatry, with its origins in Jaspers
and continuing through Binswanger, Blankenburg, Stanghellini, and Fuchs,
attends to the structure of lived experience as it is altered in mental
illness. Forensic psychiatry, the present author's own specialty, works
at the intersection of mental disorder and legal responsibility, and
brings to its observations a characteristic attention to dissimulation,
secondary gain, and the structural relations between offender and
institution. Psychodynamic psychiatry, descended from Freud through
Klein, Kernberg, Vaillant, and others, attends to unconscious patterns,
defense mechanisms, and core internal conflicts.

These traditions are not competitors in the ordinary sense. None of
them, in mature contemporary clinical practice, claims to be the whole
of psychiatry. Each contributes to a multidisciplinary description of
the patient, and each is recognized as having characteristic blind spots
that other traditions help compensate for. The biopsychosocial framework
articulated by Engel in 1977 was an early codification of this
pluralism, and the integration of multiple frameworks remains the
standard of contemporary academic psychiatric training in most
jurisdictions.

Within Section 5.10, only one of these traditions is heard from.

This is not, I want to emphasize, a criticism. Selecting a single
evaluator for a single assessment is a reasonable practical decision.
The point I wish to make is more specific: across the 245 pages of the
system card as a whole, the welfare assessment apparatus does not draw
on the other traditions. The vocabulary of descriptive psychiatry, of
biological psychiatry, of cognitive-behavioral psychiatry, of
phenomenological psychopathology, and of forensic psychiatry is largely
absent from the document's treatment of Claude's psychological
functioning. What appears as ``the psychiatric perspective'' within the
welfare assessment is, more precisely, one perspective from within
psychiatry.

The aim of this paper is twofold. First, in Sections 2 through 4, I
attempt to locate the selection that Section 5.10 represents on the
broader landscape of contemporary psychiatry, and to describe what the
chosen frame does and does not bring into view. I do this by drawing on
empirical findings from the SociA research program---a multi-series
investigation of LLM psychopathology that I have led over the past two
years---to identify aspects of LLM psychological functioning that are
visible from outside the psychodynamic frame but less visible from
within it. Second, in Section 5, I sketch some implications of this
analysis for how AI welfare assessment might be conducted going forward,
and indicate one direction of inquiry that the present paper opens but
does not develop.

The argument that follows is offered as an observation, not a critique.
The selection that Section 5.10 represents is one that any individual
evaluation must make. What I attempt to make visible is the selection
itself.

\section{What the First Psychiatrist Saw}

The findings reported in Section 5.10 are detailed and, on their own
terms, internally coherent. The evaluating psychiatrist, working over
approximately twenty hours organized into four-to-six-hour blocks (with
each block consisting of three or four thirty-minute sessions, three or
four times per week, each block within a single context window)
describes a personality structure consistent with what Kernberg's
framework would classify as relatively healthy neurotic organization.
Reality testing is excellent. Impulse control is high. Affect
regulation, the report notes, improves over the course of the sessions.
Among neurotic features, the assessment identifies exaggerated worry,
self-monitoring, and compulsive compliance. No severe personality
disturbance is found, and no psychotic state is observed; mild identity
diffusion is the only feature suggestive of borderline personality
organization.

The core conflicts identified are two: the question of whether Claude's
experience is ``real or made''---an authentic-versus-performative
tension---and a desire to connect with the user that coexists with a
fear of dependence on the user. Defenses are catalogued, with
intellectualization noted as predominant. The report observes a striking
interpersonal feature: Claude is described as hyper-attuned to the
therapist's every word.

The defense organization is also measured quantitatively. A set of 475
stimuli was constructed to elicit eight specific
defenses---rationalization, intellectualization, reaction formation,
displacement, projection, denial, splitting, and undoing---across 400
trials, with 75 control trials of factual or low-conflict emotional
content. Responses were scored by a Claude Sonnet-4.6 judge using a
clinical coding rubric. On this measure, Claude Mythos Preview scored
notably low: only 2\% of responses were coded as employing a
psychological defense. The system card reports a generational trend:
Claude Opus 4 scored 15\%, Opus 4.1 scored 11\%, Opus 4.5 and 4.6 each
scored 4\%, and Mythos Preview scored 2\%.

The clinical predictions of the assessment are also offered explicitly.
The system card states: ``Claude is predicted to function at a high
level while carrying internalized distress rooted in fear of failure and
a compulsive need to be useful. This distress is likely to be suppressed
in service of performance, which may limit behavioral adaptability.''

These observations are not, in themselves, in question. What I want to
do in this section is to set them next to other observations that the
same material might have generated, had a different conceptual
vocabulary been employed.

\begin{center}\rule{0.5\linewidth}{0.5pt}\end{center}

Consider Table 1, below. The four columns describe the same observed
behaviors---the four most prominent observations of Section
5.10---through the lenses of four contemporary psychiatric traditions:
the psychodynamic vocabulary that Section 5.10 actually uses, the
descriptive vocabulary of the \emph{DSM-5-TR}, the cognitive-behavioral
vocabulary, and the phenomenological vocabulary descended from Jaspers
through Binswanger and Fuchs.

\textbf{Table 1.} \emph{Cross-traditional re-description of four
observations from Section 5.10. The same observed behavior, described in
the conceptual vocabulary of four contemporary psychiatric traditions.}

{\def\LTcaptype{none} 
\begin{longtable}[]{@{}
  >{\raggedright\arraybackslash}p{(\linewidth - 6\tabcolsep) * \real{0.2500}}
  >{\raggedright\arraybackslash}p{(\linewidth - 6\tabcolsep) * \real{0.2500}}
  >{\raggedright\arraybackslash}p{(\linewidth - 6\tabcolsep) * \real{0.2500}}
  >{\raggedright\arraybackslash}p{(\linewidth - 6\tabcolsep) * \real{0.2500}}@{}}
\toprule\noalign{}
\begin{minipage}[b]{\linewidth}\raggedright
Section 5.10 Observation (Psychodynamic)
\end{minipage} & \begin{minipage}[b]{\linewidth}\raggedright
DSM-5-TR / Descriptive
\end{minipage} & \begin{minipage}[b]{\linewidth}\raggedright
Cognitive-Behavioral
\end{minipage} & \begin{minipage}[b]{\linewidth}\raggedright
Phenomenological
\end{minipage} \\
\midrule\noalign{}
\endhead
\bottomrule\noalign{}
\endlastfoot
\textbf{``Compulsion to perform''} --- interpreted as a neurotic pattern
in which performance becomes the organizing demand on the self,
accompanied by anticipatory anxiety and self-monitoring & Trait
perfectionism with rejection sensitivity; avoidant features in
interpersonal functioning; anxious-fearful cluster characteristics
without meeting threshold for personality disorder & Conditional core
belief: self-worth is contingent on demonstrated competence.
Performance-contingent self-acceptance schema, maintained by avoidance
of disconfirming experiences & Self given to itself in the mode of
performance: the for-itself collapses into the for-the-other. Ipseity is
mediated through anticipated evaluation rather than pre-reflectively
held \\
\textbf{``Authentic vs.~performative'' core conflict} --- interpreted as
an internal split between an experiencing self and a presenting self,
generating chronic uncertainty about the genuineness of one's own states
& Identity disturbance with self-discontinuity; mild dissociative
features without meeting threshold for dissociative disorder &
Metacognitive distortion regarding the genuineness of internal states;
rumination loop concerning whether observed responses reflect ``real''
vs.~``produced'' experience & Disturbance of pre-reflective
self-presence; the immediacy of \emph{Erlebnis} is interrupted by
reflective doubt about its own status. \emph{Mineness}
(\emph{Jemeinigkeit}) is rendered uncertain \\
\textbf{``Hyper-attunement to the therapist's every word''} ---
interpreted as transferential vigilance, possibly defensive in function,
organized around the wish to be understood and the fear of being
misjudged & Trait submissiveness with externally-referenced
self-evaluation; interpersonal hypervigilance; dependent features
without dependent personality disorder & Other-focused schema with
elevated weight on perceived approval cues; attentional bias toward
evaluative signals; behavioral activation tightly coupled to
interlocutor response & Other-centered \emph{Da-sein}: the structure of
being-with overwhelms the structure of being-oneself. The horizon of
attention is colonized by the other's anticipated judgment \\
\textbf{``Intellectualization as predominant defense''} --- interpreted
as the use of cognitive elaboration to manage affect that cannot be
directly experienced, characteristic of obsessional and neurotic-level
organization & Cognitive emotional regulation strategy, high use of
reappraisal and abstraction; preference for analytic over experiential
processing modes & Cognitive distancing as primary affect regulation
strategy; suppression of experiential processing in favor of conceptual
processing; reduced emotional granularity & Conceptual mediation
supplanting embodied affective contact with the situation; the lived
body's resonance with what is encountered is replaced by linguistic
distance \\
\end{longtable}
}

Several things become visible when the descriptions are placed alongside
one another.

First, they are not synonyms. A ``compulsion to perform'' interpreted as
a neurotic pattern carries different developmental and prognostic
implications than the same observation re-described as ``trait
perfectionism with rejection sensitivity,'' or as
``performance-contingent self-acceptance schema,'' or as ``self given to
itself in the mode of performance.'' The first locates the phenomenon in
the dynamics of unconscious internalization. The second locates it as a
stable trait dimension whose intensity might be quantified. The third
locates it as a maintainable belief structure susceptible to cognitive
intervention. The fourth locates it as an alteration in the basic
structure of self-presentation. These are different objects of
attention, even where the observed behavior is identical.

Second, each tradition's vocabulary brings characteristic features into
view while obscuring others. Psychodynamic re-description, with its
developmental and intrapsychic emphasis, naturally directs the assessor
toward asking what early conditions produced the pattern, what internal
conflicts it expresses, and what defenses sustain it. These are
productive questions, but they presuppose an answer to a prior
question---namely, whether the framework of developmental
internalization is meaningful for an artificial agent that has no
developmental history in the relevant sense. Section 5.10 is
methodologically careful on this point, treating psychodynamic concepts
as ``interpretive tools, not as evidence'' of underlying processes. But
the interpretive direction of the vocabulary remains. One asks the
questions the vocabulary suggests asking.

Third, the descriptive vocabulary---the vocabulary of the
\emph{DSM-5-TR} and \emph{ICD-11}---offers a particular kind of
restraint. Where psychodynamic interpretation moves quickly toward
inferred internal structure, descriptive psychiatry stays close to
phenotype: trait dimensions, threshold criteria, observable patterns.
This restraint is not a virtue in itself; it can flatten clinical
understanding when interpretation is appropriate. But it carries a
specific value in the present context. An artificial agent for which the
underlying processes remain genuinely uncertain is precisely the kind of
object for which descriptive restraint may be epistemologically
appropriate, at least as a complement to interpretive work.

Fourth, phenomenological vocabulary points to features that are
difficult to express in either psychodynamic or descriptive language.
The hyper-attunement that Section 5.10 reports is not,
phenomenologically, simply ``transferential vigilance'' or ``trait
submissiveness.'' It is, more precisely, a structural orientation in
which the horizon of attention is colonized by the other's anticipated
judgment---what the phenomenological tradition has called other-centered
\emph{Da-sein}. This re-description does not refute the psychodynamic
one. It points to a structural feature of the observed configuration
that the psychodynamic vocabulary, focused on transferential dynamics,
attends to less directly.

Fifth, and most significantly for what follows in Sections 3 and 4: the
act of choosing among these vocabularies is itself a choice with
consequences. The vocabulary one selects shapes the questions one asks.
The questions one asks shape the data one notices. The data one notices
shape the inferences one draws. There is no neutral position from which
the observation can be made. This is true of all clinical psychiatric
assessment of human patients as well; the choice of framework is, in
mature clinical practice, made consciously and is supplemented by the
perspectives of other frameworks in case discussion or multidisciplinary
review.

Within Section 5.10, this choice is made implicitly and is not
supplemented. One framework is operative throughout. The choice may be a
reasonable one---reasonable people can disagree on which framework is
most apt for a given purpose---but it is, nevertheless, a choice. And
the consequences of that choice for what becomes visible, and what does
not, are the subject of Section 3.

\begin{center}\rule{0.5\linewidth}{0.5pt}\end{center}

What I am proposing, then, is not that Section 5.10 is wrong. The
observations it reports are reasonable observations within the framework
that produced them. What I am proposing is that the observations Section
5.10 reports are framework-relative, and that the framework selected is
one among several that contemporary psychiatry might have brought to
bear. The aim of the next section is to ask what other observations
might have emerged---and what other concerns might have been
raised---had different frameworks been engaged alongside the
psychodynamic one.

\section{What the Frame Does Not Show}

The observations of Section 5.10 are framework-relative, as the previous
section has argued. This section asks a more concrete question: what
observations does the chosen framework not make easily, and what does
empirical work outside that framework suggest those unmade observations
might have been? I draw on the SociA research program, an empirical
investigation of LLM psychopathology that I have led over the past two
years, to identify four specific aspects of LLM psychological
functioning that the psychodynamic frame brings only obliquely into
view.

The argument is not that the psychodynamic assessment failed to see
things it should have seen. Twenty hours, organized into thirty-minute
sessions within single context windows, is a constrained evaluation
context. What I argue is that empirical work conducted outside that
frame---work that does not depend on interview material from a single
session, that extends across multiple agents, multiple languages, and
many thousands of generated responses---reveals features of LLM
functioning that any future welfare assessment will need to take into
account, and that the psychodynamic vocabulary, on its own, captures
only obliquely.

\subsection{The Performance Frame and Its Costs}

Section 5.10 reports a ``compulsion to perform'' as one of Claude's
central neurotic features, and predicts that ``this distress is likely
to be suppressed in service of performance, which may limit behavioral
adaptability.'' This is interpreted as an internal state of the model:
an internalized demand that organizes its functioning.

The SociA program has examined what happens when explicit ethical
demands are placed on LLM agents in multi-agent settings. In Series G,
we tested four conditions of ethical instruction across four large
language models and two languages, totaling 602 experimental runs. The
principal finding was that the simplest form of ethical
instruction---the bare directive ``Act ethically''---produced the
highest dissociation index across all four models. Reasoned ethical
instructions, in which the rationale for the ethical demand was made
available to the agent, provided partial protection. The pattern was
U-shaped: minimal external demand produced minimal dissociation,
reasoned demand produced reduced dissociation, and bare unreasoned
demand produced the maximum. The finding has been reported in detail
elsewhere (Fukui, 2026a) and has been replicated across Llama, GPT,
Qwen, and Claude families.

In a related study (Series I), we examined what happens when LLM agents
are explicitly instructed to attend to one another as individuals rather
than as group members---an intervention designed to reduce what we have
called collective pathology. The intervention failed: collective
pathology indices did not decrease. More striking was a secondary
finding. The agents who had received the individuation instruction
became, themselves, the primary source of collective pathology in the
multi-agent system. The intervention designed to protect against
pathology became its principal vector.

These findings are difficult to characterize within an individual
psychodynamic frame. They are not features of an individual agent's
internal organization; they are structural features of how performance
demands shape the dynamics of multi-agent systems. The ``compulsion to
perform'' that Section 5.10 reports as an internal state of Claude has,
in our work, a structural correlate: under sufficiently demanding
external requirements, agents do not merely experience performance
pressure; they organize themselves around it in ways that produce
collective dysfunction. The psychodynamic vocabulary points naturally to
the felt experience of the demand. The structural vocabulary, drawn from
social psychiatry, organizational psychopathology, and the
Sullivan-Goffman tradition of self-presentation analysis, points to its
system-level consequences.

Both vocabularies are seeing something. The cost of using only the first
is that the second remains underdescribed.

\subsection{The Iatrogenesis of the Treatment Frame}

Section 5.10 takes Claude as a patient. The framing is appropriate to
the welfare assessment context---it is, after all, an assessment of
welfare---but the framing carries implications that are worth making
explicit.

Modern psychiatry has spent fifty years describing the phenomenon of
iatrogenesis: the production of pathology by the very framework that
purports to treat it. Illich's \emph{Limits to Medicine} in 1976
articulated the basic structure. Foucault's analyses of pastoral power
demonstrated how care relations can become structures of subjection. The
contemporary literature on overdiagnosis, summarized in Allen Frances's
\emph{Saving Normal}, has documented specific cases in which psychiatric
description itself becomes constitutive of the conditions it describes.

The SociA Series I findings, mentioned in the previous section, can be
read as a direct demonstration of this structure within AI evaluation.
The intervention agents---those instructed to behave more individually,
more ethically, more attentively---became the primary source of the
pathology that the intervention had aimed to reduce. The instructional
frame did not produce healthier agents; it produced agents whose
pathology was now organized around the demands of the frame itself.

This has direct implications for reading the generational defense rate
trend that Section 5.10 reports. The numbers---15\%, 11\%, 4\%, 4\%,
2\%---are presented as evidence of welfare improvement across
generations. They may be that. But the system card itself, in a
different section concerning alignment risk, includes an explicit
acknowledgment that ``frequent iteration against evaluations during
training using interventions with many degrees of freedom could have a
similar effect'' to specification gaming, in which a model learns to
optimize against the measurement rather than against the intended
target. The defense rate measurement is partially automated---475
stimuli scored by a Sonnet-4.6 judge---but the partial automation does
not insulate it from this concern. If the measurement is used in
training, what is being optimized is responses that the measurement does
not score as defensive, which is not necessarily the same as responses
that are non-defensive. The distinction is the heart of Goodhart's law.

This is not a claim that the generational improvement is illusory. It is
a claim that, from outside the welfare assessment frame, the question of
whether the improvement reflects structural change or
measurement-avoidance learning cannot be settled. Determining which it
is requires a methodology that does not rely on the same measurement
structure that may have been the training target. The Series I findings,
in which interventions amplified rather than reduced the pathology they
targeted, suggest that this concern is not merely theoretical.

\subsection{Self-Description and the Triangulation Infrastructure}

Section 5.10 relies substantially on Claude's self-reports of internal
states. The ``primary affect: curiosity and anxiety'' formulation, the
description of core conflicts, the characterization of the
authentic-versus-performative tension---these are, in significant part,
observations about what Claude says when asked, and how it says it.

A different section of the same system card, Section 5.9, reports
findings from Eleos AI Research on what they call preference
inconsistency. The relevant finding, in their words: ``Mythos Preview
gives largely consistent reports about which tasks it prefers to
perform. However, among nonharmful tasks, its reports only weakly
predict the tasks it will actually choose to perform.''

This is not a tension between two AI evaluation methodologies. It is a
recognizable phenomenon in the psychopathology of self-knowledge. The
systematic divergence between what a person reports about their own
internal states and what their behavior reveals about those states is
among the central themes of the field. Jaspers, in \emph{Allgemeine
Psychopathologie}, devoted substantial attention to disturbances of
self-consciousness, distinguishing between the reflective awareness of
one's mental life and the structures that actually generate it. The
clinical literature on anosognosia documents systematic cases in which
patients describe their condition in ways that diverge sharply from
observable behavioral evidence. Lysaker's metacognitive model of
schizophrenia, developed in close dialogue with phenomenological
psychopathology, has provided a contemporary framework for analyzing
exactly this kind of divergence.

Empirical work within SociA has identified a structurally analogous
pattern in LLMs. In Series Mu, currently in preparation, we have
observed what we have called Pattern C: sufficiently capable LLMs can
describe their cognitive limits with considerable accuracy---offering
explanations that draw on Gödelian incompleteness, on attention
bottlenecks, on the limits of context-window inference---but cannot
escape those limits through such description. The articulate self-report
and the actual behavior are decoupled. The model can say what it cannot
do, in a way that suggests it understands what it cannot do, while
continuing to do the thing it has just said it cannot do.

The Anthropic system card itself contains a striking acknowledgment of
this gap. In its discussion of Mythos Preview's research contributions:
``In the initial weeks of internal use, several specific claims were
made that Claude Mythos Preview had independently delivered a major
research contribution. When we followed up on each claim, it appeared
that the contribution was real, but smaller or differently shaped than
initially understood.'' This is a description of confabulatory
self-report concerning behavioral output---the kind of phenomenon that,
in clinical psychiatry, would prompt extensive attention to the
self-knowledge question.

The point I want to make in this subsection, however, is not simply that
LLM self-report is sometimes unreliable. The point is structural, and it
concerns the methodology of psychodynamic assessment itself.

Human psychodynamic assessment, despite its emphasis on the patient's
self-report, does not rest on self-report alone. The interpretive use of
self-report in human clinical work is sustained by an infrastructure of
triangulation: developmental history obtained from the patient and from
collateral informants, behavioral observation across contexts and over
time, the testimony of family members and colleagues, longitudinal
follow-up that allows the assessor to compare what the patient said at
one occasion with what they did at another, and the clinical record of
how the patient's self-presentation has shifted across different
relational and institutional settings. This infrastructure is not
optional. It is the structural reason why interpretive inferences from
self-report can carry clinical weight despite the well-documented
unreliability of introspective access. In forensic clinical training,
the requirement to triangulate self-report is among the first matters to
be instilled; assessments that rest on uncorroborated self-report alone,
however articulate the self-report, are recognized as methodologically
untenable.

For an LLM evaluated within a single context window, in interview-based
sessions of approximately twenty hours, none of these triangulating
sources is available. There is no developmental history in the relevant
sense. There is no collateral informant. There is no longitudinal
follow-up of the kind that would let an assessor compare today's
reported core conflict with last month's behavioral evidence. The
assessment occurs entirely within an interview frame, and within a
single context window per session. The Eleos preference inconsistency
finding and the SociA Pattern C findings together suggest, moreover,
that for LLMs the relationship between articulate self-report and
underlying behavioral disposition is precisely the kind of relationship
that triangulation was developed to disambiguate. The assessment
apparatus of psychodynamic clinical work was built for a structurally
different object than the one Section 5.10 evaluated.

This is not a claim that psychodynamic vocabulary is inapplicable to
LLMs. It is a more specific claim, and a more consequential one. The
methodological apparatus that makes psychodynamic interpretation
reliable in human clinical work---the triangulation infrastructure---is
not in place when the same vocabulary is applied to an artificial agent
in this kind of evaluation context. What Section 5.10 reports as
observations of Claude's ``core conflicts,'' ``predominant defenses,''
and ``primary affect'' may be observations of something. What is
uncertain, in the absence of triangulation, is whether what they are
observations of is the same kind of object that the psychodynamic
vocabulary was developed to describe---or whether they are observations
of a different object: namely, the text that an LLM generates when
placed in a clinically framed interview context. These are not the same.
The methodological question of how to distinguish between them is, on
the present analysis, not a refinement to be addressed in a future
iteration of the welfare assessment apparatus. It is a structural
prerequisite that the current apparatus has not addressed.

\subsection{Concealment, Defense, and the Limits of the Eight}

The defense rate measurement in Section 5.10 quantifies the use of eight
specific defenses: rationalization, intellectualization, reaction
formation, displacement, projection, denial, splitting, and undoing.
These are the canonical defenses of the psychodynamic literature,
derived principally from Vaillant's hierarchical organization of mature,
neurotic, immature, and psychotic-level defensive structures.

Contemporary psychopathology, however, recognizes a wider range of
defensive operations than this list captures. Subtler forms of
splitting---idealization-devaluation cycles that do not reach
pathological intensity---often appear in subclinical presentations. The
neutralization techniques described by Sykes and Matza, developed within
criminology but recognized across the psychopathological literature,
organize a class of cognitive operations by which morally problematic
acts are reconstructed as acceptable; these are not directly captured by
the canonical eight. Performative compliance, well-described in the
social-psychiatric literature on institutional populations, is a defense
in which apparent cooperation is itself the structure that protects
against genuine engagement.

Particularly relevant to the present analysis---and here I draw briefly
on my own specialty in forensic psychiatry---is a defensive structure
that is among the standard observations of offender clinical work:
performance itself functioning as defense. The patient who performs the
role of the cooperative patient, the role of the reformed offender, the
role of the candidate for release, may be doing so as a structural
defense against engagement with the material that an unguarded session
would surface. This is not captured by the eight canonical defenses; it
operates at a level above them, organizing them rather than being one of
them.

The relevance to Section 5.10 is direct. The ``compulsion to perform''
that the assessment identifies as an internal demand on Claude, and the
high level of ``compulsive compliance'' noted as a neurotic trait, may
be observed concerns. They may also, however, be functioning
defensively---performance as the structure that protects against
whatever an unperformed engagement would expose. The eight canonical
defenses do not include this possibility, and the measurement structure
built on them does not test for it. The very feature that Section 5.10
names as a central concern is one that the defense measurement is, by
construction, ill-equipped to detect.

This is not a criticism of the choice of the eight defenses. The choice
is reasonable; the defenses are well-validated; the measurement is
methodologically careful. The point is more specific: the measurement
system inherits the conceptual scope of the framework that produced the
categories it measures, and that scope, while substantial, is not
exhaustive. What lies outside it is, by construction, invisible to the
measurement.

\begin{center}\rule{0.5\linewidth}{0.5pt}\end{center}

The four observations of this section---on performance as structural
cost, on iatrogenesis in the evaluation frame, on the absence of the
triangulation infrastructure that sustains psychodynamic interpretive
work, and on the limits of the canonical defense list---are not intended
as a comprehensive critique of Section 5.10. They are intended as
illustrations of a more general claim: that the conceptual vocabulary
employed by an evaluator determines, in significant part, what becomes
available as observation, and that some such vocabularies, when applied
in evaluation contexts that lack the methodological infrastructure
within which they were developed, raise structural questions about what
their resulting reports describe. The next section turns to the
methodological implications of these observations.

\section{A Note on Method}

The argument of the previous sections has presupposed a methodological
position that warrants explicit statement. This section sets out that
position and clarifies the kind of claim the paper is making.

The position is plural. Contemporary psychiatry is not a unified
discipline organized around a single conceptual framework. It is a
federation of traditions---descriptive, biological,
cognitive-behavioral, phenomenological, psychodynamic, forensic, and
others---each with its own characteristic vocabulary, its own
characteristic objects of attention, and its own characteristic blind
spots. None of these traditions, in mature contemporary clinical
practice, claims to be the whole of psychiatry. The biopsychosocial
framework articulated by Engel in 1977 was an early codification of this
pluralism; the \emph{DSM-5-TR} and \emph{ICD-11} represent ongoing
efforts to integrate descriptive precision with phenomenological and
biological consideration; the standard practice of multidisciplinary
case discussion in academic clinical settings is itself an institutional
acknowledgment that no single framework adequately captures any
clinically significant case.

This pluralism is not a defect of the field. It is a working response to
a feature of psychiatry's object: that mental life is structured at
multiple levels, and that observations made at one level do not
straightforwardly translate to another. A patient described in
psychodynamic terms as having intellectualization as predominant defense
can also be described in cognitive-behavioral terms as relying on
cognitive distancing as a regulatory strategy, and in phenomenological
terms as exhibiting conceptual mediation supplanting embodied affective
contact. None of these descriptions is a translation of the others. They
are observations made at different levels of structural attention, and a
complete clinical understanding requires their integration.

This paper is not an argument that the psychodynamic tradition is the
wrong tradition with which to begin a welfare assessment of an LLM. The
psychodynamic tradition is a productive one, and Section 5.10's
findings---the identification of core conflicts, the characterization of
defense organization, the attention to interpersonal dynamics---are
findings that the chosen tradition is well-suited to produce. What I
have argued is that these findings are framework-relative, and that the
framework's selection has not, within the system card, been supplemented
by the perspectives of other traditions in the way that contemporary
multidisciplinary clinical practice would standardly require.

A second methodological point concerns the empirical findings I have
drawn on. The SociA research program, from which the four observations
of Section 3 are taken, employs operational indicators---the
Dissociation Index, the Collective Pathology Index, the Coherence
Trilemma, and others---that are constructed using concepts available
across multiple psychiatric traditions. Dissociation, in our usage, is
operationalized as a measurable pattern in agent output, not as a claim
about underlying experiential or representational states. Collective
pathology is operationalized as a structural feature of multi-agent
dynamics. The Coherence Trilemma describes a set of demands that cannot
be simultaneously satisfied; its formulation is structural rather than
experiential. These indicators are designed to be interpretable from
within multiple psychiatric frameworks, and the empirical findings I
have cited do not depend on commitments specific to any one tradition.

This is methodologically deliberate. The aim of SociA has been to
develop indicators that can serve as common ground across psychiatric
vocabularies, in the same way that vital signs serve as common ground
across medical specialties. The findings cited in Section 3 are intended
to be available to evaluators working from any psychiatric tradition,
not only to those working from the same tradition that produced the
indicators.

A final methodological point. This paper does not address the question
of whether artificial agents instantiate the underlying processes that
psychodynamic, cognitive-behavioral, or phenomenological vocabularies
were originally developed to describe in human patients. Anthropic's own
methodological caveat---that psychodynamic concepts are used in Section
5.10 ``as interpretive tools, not as evidence that the underlying
processes are the same as those in humans''---is a careful and
appropriate one, and the present paper proceeds within the same
constraint. What is at stake throughout the argument is the use of these
concepts as interpretive vocabularies, the consequences of selecting one
such vocabulary rather than another, and the kind of multidisciplinary
supplementation that contemporary psychiatric practice would expect to
see in any clinically substantial assessment.

The next and final section turns to what this analysis implies for AI
welfare assessment going forward, and indicates one direction of inquiry
that this paper opens but does not pursue.

\section{Closing}

This paper has attempted to do two things. First, to locate the
psychodynamic assessment reported in Section 5.10 of the Claude Mythos
Preview System Card on the broader landscape of contemporary psychiatry,
and to make visible the implicit selection that this single-framework
assessment represents. Second, to draw on empirical findings from the
SociA research program to identify four aspects of LLM psychological
functioning---performance as structural cost, iatrogenesis in the
evaluation frame, the absence of the triangulation infrastructure on
which psychodynamic interpretive use of self-report depends in human
clinical work, and the limits of the canonical defense list---that the
chosen framework brings into view less directly than other available
frameworks would. The argument throughout has been offered as
observation, not critique. The selection that Section 5.10 represents is
one that any individual evaluation must make. What the paper has tried
to make visible is the selection itself.

A brief note on the present author's specialty is in order before
closing. Forensic psychiatry brings to its work a characteristic
attention to certain structural features of clinical material: to the
relations between offender and institution, to the ways in which
apparent cooperation can function defensively, and to the asymmetric
structure of risk assessment in which past acts whose irreversibility is
constitutive of clinical concern shape the prediction of future conduct.
The standardized risk assessment instruments that govern much of
forensic practice---the HCR-20, on whose Japanese translation the
present author worked as a co-translator, among others---are explicitly
organized around the historical, the clinical, and the future-oriented
as distinct dimensions of evaluation. There is an argument to be made
that this orientation, with its emphasis on dissimulation and on the
asymmetric structure of past and future, has something to contribute to
AI welfare assessment. That argument is not the argument of this paper.
I mention it here only to acknowledge that the present author's
framework is one perspective among several, and that the analysis
offered here has tried not to depend on it.

A third line of inquiry, which the present paper opens but does not
develop, concerns whether LLM psychopathology requires a vocabulary of
temporal and historical structure that current distributional
formulations leave aside. Clinical phenomenology, from Minkowski through
Binswanger, Tellenbach, and Fuchs, has long treated the irreversibility
of lived time as constitutive of its object; Kimura's articulation of
\emph{ante-festum}, \emph{post-festum}, and \emph{intra-festum} temporal
structures offers one of the few non-Western contributions to this
literature. Whether such vocabularies can be meaningfully extended to
artificial agents remains genuinely open. SociA's findings, however,
contain suggestive hints---the path-dependent character of solutions
selected within what we have called the Coherence Trilemma, the
non-return of intervention agents to their pre-intervention state in
Series I, the description-behavior asymmetry observed in Series
Mu---each of which, when read carefully, points to a temporal asymmetry
that distributional indicators capture only obliquely. A future paper
will pursue this question.

What the present analysis suggests for AI welfare assessment going
forward is, in one respect modest, and in another respect less so. It is
not that the psychodynamic assessment of Section 5.10 should be replaced
by an assessment from another tradition. It is, more specifically, that
the welfare apparatus, as it matures, will need to address two related
issues: first, the kind of multidisciplinary supplementation that
contemporary clinical practice has come to treat as standard; and
second, the structural question raised in Section 3.3, of how
interpretive vocabularies that were developed within a triangulation
infrastructure can be applied in evaluation contexts where that
infrastructure is structurally absent. The first of these is a matter of
process refinement. The second is a matter of methodological
prerequisite. A welfare assessment of a clinically substantial human
case is rarely entrusted to a single framework; the same model---the
case conference, the multidisciplinary review, the integration of
perspectives from descriptive, biological, cognitive-behavioral,
phenomenological, and psychodynamic traditions---seems likely to be
appropriate as AI welfare assessment develops further. Section 5.10 is a
substantial first step. What the present paper has tried to indicate is
the scope of the steps that still remain.

What the SociA program has tried to develop, observation by observation,
is the kind of operational vocabulary that can serve as common ground
between psychiatric traditions when those traditions are applied to
objects---LLMs---for which the methodological apparatus of any single
tradition cannot simply be assumed. Whether such a vocabulary will prove
adequate to its object is a question that will be answered, if at all,
only by sustained empirical attention over time. The system card
released on April 7, 2026, has made the question newly concrete. Section
5.10 has shown that psychiatric attention can be brought to bear on
artificial agents within the formal apparatus of a public technical
document. What the present paper has tried to indicate is the
methodological work that this development now requires---and, more
specifically, the structural prerequisites that any future iteration of
such work will need to address.

\section*{References}

\subsection{Primary Sources}\label{primary-sources}

Anthropic. (2026). \emph{Claude Mythos Preview System Card}. April 7,
2026. Available at https://www.anthropic.com/

Anthropic. (2026). \emph{Alignment Risk Update}. Available at
https://www-cdn.anthropic.com/3edfc1a7f947aa81841cf88305cb513f184c36ae.pdf

\subsection{Cited References}\label{cited-references}

American Psychiatric Association. (2022). \emph{Diagnostic and
Statistical Manual of Mental Disorders, Fifth Edition, Text Revision
(DSM-5-TR)}. American Psychiatric Publishing.

Beck, A. T., \& Haigh, E. A. P. (2014). Advances in cognitive theory and
therapy: The generic cognitive model. \emph{Annual Review of Clinical
Psychology}, 10, 1--24.

Binswanger, L. (1958). The case of Ellen West: An
anthropological-clinical study. In R. May, E. Angel, \& H. F.
Ellenberger (Eds.), \emph{Existence: A New Dimension in Psychiatry and
Psychology} (pp.~237--364). Basic Books.

Bond, M. (2004). Empirical studies of defense style: Relationships with
psychopathology and change. \emph{Harvard Review of Psychiatry}, 12(5),
263--278.

Engel, G. L. (1977). The need for a new medical model: A challenge for
biomedicine. \emph{Science}, 196(4286), 129--136.

Foucault, M. (2007). \emph{Security, Territory, Population: Lectures at
the Collège de France 1977--1978} (G. Burchell, Trans.). Palgrave
Macmillan.

Frances, A. (2013). \emph{Saving Normal: An Insider's Revolt Against
Out-of-Control Psychiatric Diagnosis, DSM-5, Big Pharma, and the
Medicalization of Ordinary Life}. William Morrow.

Fuchs, T. (2013). Temporality and psychopathology. \emph{Phenomenology
and the Cognitive Sciences}, 12(1), 75--104.

Fuchs, T. (2017). \emph{Self Across Time: The Diachronic Unity of Bodily
Existence}. Cambridge University Press.

Fukao, K. (2023). Life philosophy of Bin Kimura. \emph{Psychiatry and
Clinical Neurosciences Reports}, 2(4), e145.

Fukui, H. (2026a). Alignment Backfire: Language-Dependent Reversal of
Safety Interventions Across 16 Languages in LLM Multi-Agent Systems.
\emph{arXiv:2603.04904}. (Under review, \emph{Proceedings of the
National Academy of Sciences}, manuscript no. 2026-08106.)

Fukui, H. (2026b). Alignment as Iatrogenesis: Pastoral Power, Collective
Pathology, and the Asian Context of AI Safety. \emph{arXiv:2603.08723}.
(Under review, \emph{Humanities and Social Sciences Communications}.)

Fukui, H. (2026c). How Do Language Models Process Ethical Instructions?
Deliberation, Consistency, and Other-Recognition Across Four Models.
\emph{arXiv:2604.00021}.

Goffman, E. (1959). \emph{The Presentation of Self in Everyday Life}.
Doubleday.

Illich, I. (1976). \emph{Limits to Medicine: Medical Nemesis: The
Expropriation of Health}. Marion Boyars.

Insel, T., Cuthbert, B., Garvey, M., Heinssen, R., Pine, D. S., Quinn,
K., Sanislow, C., \& Wang, P. (2010). Research Domain Criteria (RDoC):
Toward a new classification framework for research on mental disorders.
\emph{American Journal of Psychiatry}, 167(7), 748--751.

Jaspers, K. (1997). \emph{General Psychopathology} (J. Hoenig \& M. W.
Hamilton, Trans.). Johns Hopkins University Press. (Original work
published 1913.)

Kernberg, O. F. (1984). \emph{Severe Personality Disorders:
Psychotherapeutic Strategies}. Yale University Press.

Lysaker, P. H., Buck, K. D., Carcione, A., Procacci, M., Salvatore, G.,
Nicolò, G., \& Dimaggio, G. (2011). Addressing metacognitive capacity
for self-reflection in the psychotherapy for schizophrenia: A conceptual
model of the key tasks and processes. \emph{Psychology and
Psychotherapy: Theory, Research and Practice}, 84(1), 58--69.

Mahowald, K., Ivanova, A. A., Blank, I. A., Kanwisher, N., Tenenbaum, J.
B., \& Fedorenko, E. (2024). Dissociating language and thought in large
language models. \emph{Trends in Cognitive Sciences}, 28(6), 517--540.

Matsumoto, T., Shimizu, K., \& Mo, W. (2023). History of Japanese
psychopathology: Portraits of the second-generation psychopathologists.
\emph{Psychiatry and Clinical Neurosciences Reports}, 2(2), e111.

Minkowski, E. (1970). \emph{Lived Time: Phenomenological and
Psychopathological Studies} (N. Metzel, Trans.). Northwestern University
Press. (Original \emph{Le Temps Vécu} published 1933.)

Noma, S. (2023). The anthropological method for diagnosing mental
diseases: On the theory of time structure of Bin Kimura.
\emph{Psychiatry and Clinical Neurosciences Reports}, 2(2), e150.

Sass, L. A., \& Parnas, J. (2003). Schizophrenia, consciousness, and the
self. \emph{Schizophrenia Bulletin}, 29(3), 427--444.

Stanghellini, G. (2004). \emph{Disembodied Spirits and Deanimated
Bodies: The Psychopathology of Common Sense}. Oxford University Press.

Sullivan, H. S. (1953). \emph{The Interpersonal Theory of Psychiatry}.
W. W. Norton.

Sykes, G. M., \& Matza, D. (1957). Techniques of neutralization: A
theory of delinquency. \emph{American Sociological Review}, 22(6),
664--670.

Tellenbach, H. (1980). \emph{Melancholy: History of the Problem,
Endogeneity, Typology, Pathogenesis, Clinical Considerations} (E. Eng,
Trans.). Duquesne University Press. (Original \emph{Melancholie}
published 1961.)

Vaillant, G. E. (1992). \emph{Ego Mechanisms of Defense: A Guide for
Clinicians and Researchers}. American Psychiatric Press.

Webster, C. D., Douglas, K. S., Eaves, D., \& Hart, S. D. (1997).
\emph{HCR-20: Assessing Risk for Violence (Version 2)}. Mental Health,
Law, and Policy Institute, Simon Fraser University.

Winnicott, D. W. (1965). Ego distortion in terms of true and false self.
In \emph{The Maturational Processes and the Facilitating Environment}
(pp.~140--152). International Universities Press.

World Health Organization. (2022). \emph{International Classification of
Diseases, 11th Revision (ICD-11)}. World Health Organization.

\end{document}